\documentclass[12pt,onecolumn]{revtex4}

\begin{document}

\title{Rotating Topological Black Brnaes in Various Dimensions\\
and AdS/CFT Correspondence}
\author{M. H. Dehghani}
\address{Physics Department and Biruni  Observatory,
         Shiraz University, Shiraz 71454, Iran}
\begin{abstract}
We consider rotating topological black branes with one rotational
parameter in various dimensions. Also a general five-dimensional
higher genus solution of the Einstein equation with a negative
cosmological constant which represents a topological black brane
with two rotational parameters is introduced. We find out that the
counterterms inspired by conformal field theory introduced by
Kraus, Larsen and Sieblink cannot remove the divergences in $r$ of
the action in more than five dimensions. We modify the
counterterms by adding a curvature invariant term to it. Using the
modified counterterms we show that the $r$ divergences of the
action, the mass, and the angular momentum densities of these
spacetimes are removed. We also find out in the limit of $m=0$ the
mass density of these spacetimes in odd dimensions is not zero.
\end{abstract}

\maketitle


\section{Introduction\label{Intro}}

The AdS conformal field theory (CFT) correspondence conjecture
asserts that there is a relation between a super gravity or string
theory in $(n+1)$-dimensional anti-de Sitter (AdS) spacetimes and
a conformal field theory (CFT) living on an $n$-dimensional
boundary \cite{Mal1}. This equivalence enables one to remove the
divergences of the action and conserved quantities of gravity in
the same way as one does in field theory. Indeed the CFT-inspired
boundary counterterms furnish a means for calculating the action
and conserved charges intrinsically without reliance on any
reference spacetime \cite{Kls,BK1,Hen}. The efficiency of this
approach has been demonstrated in a broad range of examples in the
spacially infinite limit, where the AdS/CFT correspondence applies
\cite{Emp,Mann,Awad}. Although the AdS/CFT correspondence applies
for the case of spacially infinite boundary, it was also employed
for the computation of the conserved and thermodynamical
quantities in the case of a finite boundary \cite{Deh1}. These
ideas have been recently extended to the case of asymptotically de
Sitter spacetimes \cite{Stro1,Deh2}.

In this paper we want to apply the AdS/CFT correspondence to the
case of rotating topological black branes whose $(n-1)$-surfaces
at fixed $r$ and $t$ have nontrivial topology with nonconstant
curvature \cite{Kl97,Kl98}. Although in the non-rotating case one
can compactify these hypersurfaces to obtain topological black
holes \cite{TBH1,TBH2}, in the rotating case this cannot be done.
The AdS/CFT correspondence has been used for the case of
topological black holes with constant curvature horizon
\cite{Emp,Cai,Ghez}, but it has not applied to the case of
rotating higher genus black branes until now. Thus it is
worthwhile to investigate the efficiency of the AdS/CFT
correspondence for this type of black branes. It is remarkable to
mention that the counterterms introduced in Ref. \cite{Kls} cannot
remove the $r$ divergences in the action of rotating topological
black branes in more than five dimensions. But as it is known the
expression for $I_{ct}$ obtained by the algorithm given in Ref.
\cite{Kls} is not unique, and one can add any functional of
curvature invariants that vanish at infinity in a given dimension.
So, we first modify the counterterms and then we show that the $r$
divergences of the action and the conserved charge densities of
these spacetimes in various dimension up to seven are removed.
Also we introduce the general five-dimensional topological
Kerr-AdS solution which describes rotating black brane in AdS
spacetime with two rotational parameters.

The outline of our paper is as follows. In Sec. \ref{Action} we review the
basic formalism and introduce the modified counterterms. Section \ref{Kerr}
will be devoted to the consideration of topological Kerr-AdS metric with one
rotational parameter in various dimensions. We also show that the modified
counterterms remove the divergences in $r$ for the densities of the action
and the conserved charges. In Sec. \ref{GenK5} the general five-dimensional
topological Kerr-AdS spacetimes with two rotational parameters is
introduced, and the action and conserved charge densities computed. We
finish our paper with some concluding remarks.

\section{The Action and Conserved Quantities\label{Action}}

The gravitational action of $(n+1)$-dimensional spacetimes
$\mathcal{M}$, with boundary $\delta \mathcal{M}$ is
\begin{equation}
I_G=-\frac 1{16\pi }\int_{\mathcal{M}}d^{n+1}x\sqrt{-g}\left( \mathcal{R}%
\text{ }-2\Lambda \right) +\frac 1{8\pi }\int_{\partial \mathcal{M}}d^nx%
\sqrt{-\gamma }\Theta (\gamma ).  \label{Actg}
\end{equation}
The first term is the Einstein-Hilbert volume (or bulk) term with
negative cosmological constant $\Lambda =-n(n-1)/(2l^2)$ and the
second term is the Gibbons-Hawking boundary term which is chosen
such that the variational principle is well-defined. The manifold
$\mathcal{M}$ has metric $g_{\mu \nu }$ and covariant derivative
$\nabla _\mu $. $\Theta $ is the trace of the extrinsic curvature
$\Theta ^{\mu \nu }$ of any boundary(ies) $\partial \mathcal{M}$
of the manifold $\mathcal{M}$, with induced metric(s) $\gamma
_{ij}$. In general the first and second terms of Eq. (\ref{Actg})
are both divergent when evaluated on solutions, as is the
Hamiltonian, and other associated conserved quantities. Rather
than eliminating these divergences by incorporating a reference
term in the spacetime \cite{BY,BCM}, a new term, $I_{ct}$, is
added to the action which is a functional only of the boundary
curvature invariants. For an asymptotically AdS spacetime, this
has been done in Ref. \cite{Kls} through the use of an algorithmic
procedure. These counterterms up to seven dimensions are
\begin{equation}
I_{ct}=\frac 1{8\pi }\int_{\partial \mathcal{M}_\infty }d^nx\sqrt{-\gamma }%
\left\{-\frac{n-1}l-\frac{lR}{2(n-2)}-\frac{l^3}{2(n-4)(n-2)^2}%
\left(R_{ab}R^{ab}-\frac n{4(n-1)}R^2\right)+...\right\},
\label{Actct}
\end{equation}
where $R$, $R_{abcd}$, and $R_{ab}$ are the Ricci scalar, Riemann
tensor and Ricci tensor of the boundary metric $\gamma _{ab}$.
Indeed, there may exist a very large number of possible invariants
one could add in a given dimension, but in Ref. \cite{Kls} only a
finite number of them have been introduced. These counterterms
have been used by many authors for a wide variety of the
spacetimes, including Schwarzschild-AdS, topological
Schwarzschild-AdS, Kerr-AdS, Taub-NUT-AdS, Taub-bolt-AdS, and
Taub-bolt-Kerr-AdS \cite {Emp,Mann,Awad}. Although the
counterterms introduced in Ref. \cite{Kls} can remove the
divergences of the action and conserved charges of these
spacetimes, one may show that these counterterms cannot remove the
divergences in $r$ of the action of topological Kerr-AdS in more
than five dimensions. In order to remove the $r$ divergences, we
should modify the counterterms (\ref{Actct}) by adding the
following curvature invariant term to it:
\begin{equation}
I_{ct}^{\prime }=\frac 1{8\pi }\int_{\partial \mathcal{M}_\infty }d^nx\sqrt{%
-\gamma }\frac{l^3}{2(n-1)(n-4)(n^2-4)}\nabla ^2R.  \label{mict}
\end{equation}
It is remarkable to mention that the counterterm (\ref{mict}) is
equal to zero for the case of black holes whose horizons have
positive or negative constant curvature. The total action can be
written as a linear combination of the gravity term (\ref{Actg})
and the counterterms (\ref{Actct}) and (\ref{mict}) as
\begin{equation}
I=I_G+I_{ct}+I_{ct}^{\prime }.  \label{totact}
\end{equation}
Using the Brown and York definition \cite{BY} one can construct a
divergence free stress-energy tensor from the total action
(\ref{totact}) as
\begin{eqnarray}
T^{ab} &=&\frac 1{8\pi } \{ (\Theta ^{ab}-\Theta \gamma ^{ab})-\frac{n-1}%
l\gamma ^{ab}+\frac l{n-2}(R^{ab}-\frac 12R\gamma ^{ab})  \nonumber \\
&&+\frac{l^3}{(n-4)(n-2)^2}[-\frac 12\gamma
^{ab}(R^{cd}R_{cd}-\frac
n{4(n-1)}R^2)-\frac{n}{2(n-2)}R R^{ab}  \nonumber \\
&&+2R_{cd}R^{acbd}-\frac{n-2}{2(n-1)}\nabla ^a\nabla ^bR+\nabla
^2R^{ab}-\frac 1{2(n-1)}\gamma ^{ab} \nabla ^2 R]+... \}.
\label{Stres}
\end{eqnarray}

To compute the conserved charges of the spacetime, one should choose a
spacelike surface $\mathcal{B}$ in $\partial \mathcal{M}$ with metric $%
\sigma _{ij}$, and write the boundary metric in ADM form:
\[
\gamma _{ab}dx^adx^a=-N^2dt^2+\sigma _{ij}\left( d\phi ^i+V^idt\right)
\left( d\phi ^j+V^jdt\right) ,
\]
where the coordinates $\phi ^i$ are the angular variables
parametrizing the hypersurface of constant $r$ around the origin.
Then the conserved quantities associated with the stress tensors
of Eq. (\ref{Stres}) can be written as
\begin{equation}
\mathcal{Q}(\mathcal{\xi )}=\int_{\mathcal{B}_\infty }d^{n-1}\phi \sqrt{%
\sigma }T_{ab}n^a\mathcal{\xi }^b,  \label{charge}
\end{equation}
where $\sigma $ is the determinant of the metric $\sigma _{ij}$, $\mathcal{%
\xi }$ and $n^a$ are the Killing vector field and the unit normal
vector on the boundary $\mathcal{B}$ .

Now in the following sections we study the implications of
including the modified counterterms for the class of rotating
topological spacetimes in various dimensions. For each Killing
vector $\mathcal{\xi }$ , there exist an associated conserved
charge. For our case, rotating topological spacetimes, the first
Killing vector is $\xi =\partial /\partial t$ and therefore its
associated conserved charge is the total mass of the system
enclosed by the boundary given by
\begin{equation}
M=\int_{\mathcal{B}_\infty }d^{n-1}\phi \sqrt{\sigma }T_{ab}n^a\xi ^b.
\label{Mas}
\end{equation}
The charge associated to a rotational Killing symmetry generated
by $\zeta =\partial /\partial \phi $ is the angular momentum
written as
\begin{equation}
J=\int_{\mathcal{B}_\infty }d^{n-1}\phi \sqrt{\sigma }T_{ab}n^a\zeta ^b,
\label{Ang}
\end{equation}
but in our case there is no global rotational Killing symmetry.
However, the vector $\zeta =\partial /\partial \phi $, although it
is not a Killing vector, obeys locally the condition $\nabla
_{(a}\zeta _{b)}=0$ and is therefore a kind of approximate
symmetry, or a locally exact symmetry.

In the context of the AdS/CFT correspondence the limit in which the boundary
$\mathcal{B}$ becomes infinite ($\mathcal{B}_\infty $) is taken, and the
counterterm prescription ensures that the divergences in $r$ of the action,
mass, and angular momentum are removed. No embedding of the surface $\mathcal{%
B}$ into a reference spacetime is required and the quantities which are
computed are intrinsic to the spacetimes. In our case since the boundary $%
\mathcal{B}$ cannot be compactified, the divergences in the
angular variables parametrizing the hypersurface $\mathcal{B}$
remain.

\section{Topological Kerr-AdS Metric With One Rotational Parameter\label
{Kerr}}

In this section we consider the class of higher genus ($g>1$)
rotating topological solutions of the Einstein equation with
negative cosmological constant in various dimensions with one
rotational parameter. These solutions in $(n+1)$ dimension can be
written as \cite{Kl97,Kl98}:
\begin{eqnarray}
ds^2 &=&-\frac{\Delta _r}{\rho ^2}(dt+\frac a\Xi \sinh ^2\chi d\phi )^2+%
\frac{\rho ^2}{\Delta _r}dr^2+\frac{\rho ^2}{\Delta _\chi }d\chi ^2
\nonumber \\
&&\ +\frac{\Delta _\chi \sinh ^2\chi }{\rho ^2}[a
dt-\frac{(r^2+a^2)}\Xi d\phi ]^2+r^2\cosh ^2\chi d\Sigma _{n-3},
\label{met1a}
\end{eqnarray}
where $d\Sigma _{n-3}$ is the metric of a unit $(n-3)$-dimensional compact
pseudosphere and
\begin{eqnarray}
\Delta _r &=&(r^2+a^2)(\frac{r^2}{l^2}-1)-2mr^{4-n},  \nonumber \\
\Delta _\chi &=&1+\frac{a^2}{l^2}\cosh ^2\chi ,  \nonumber \\
\Xi &=&1+\frac{a^2}{l^2},  \nonumber \\
\rho ^2 &=&r^2+a^2\cosh ^2\chi .  \label{met1b}
\end{eqnarray}
One may note that Eqs. (\ref{met1a} and \ref{met1b}) describe
spacetimes which reduce in the case of $a=0$, to the topological
Schwarzschild -AdS black holes in $(n+1)$ dimension. Hence we
expect the parameters $m$ and $a$ to be associated with the mass
and angular momentum of the spacetime respectively. The metric of
Eq. (\ref{met1a}) is a limit case of the generalized Petrov-type D
solution of the Einstein equation with cosmological constant in
$(n+1)$ dimension introduced in \cite{Kl98}.

The metric induced on a spacelike $(n-1)$-surface at fixed
coordinates $r$ and $t$ is
\begin{equation}
\frac{\rho ^2}{\Delta _\chi }d\chi ^2+[(r^2+a^2)^2\Delta _\chi -a^2\sinh
^2\chi \Delta _r]\frac{\sinh ^2\chi }{\rho ^2}d\phi ^2+r^2d\Sigma _{n-3}.
\label{surmet}
\end{equation}
Note that the Gaussian curvature of this metric is not constant
and one cannot compactify the $(\chi -\phi )$ sector \cite{Kl98}.
In order to have a Euclidean metric we have to require that the
term in the bracket of Eq. (\ref {surmet}) be positive. That is
\begin{equation}
(r^2+a^2)^2\Delta _\chi -a^2\sinh ^2\chi \Delta _r>0.  \label{Eucond}
\end{equation}
This condition restricts the allowed values of the mass parameter
and we will consider it in various dimensions in the following
sections.

\subsection{Topological Kerr-AdS$_4$ Metric}

\ The metric of Eqs. (\ref{met1a}) and (\ref{met1b}) for $n=3$ is
a limit case of the metric of Plebanski and Demianski \cite{Pleb}
which is the most general known Petrov type-D solution of the
source-free Einstein-Maxwell equation with cosmological constant
in four dimensions \cite{Kl97}. This metric has two inner and
outer horizons located at $r_{-}$ and $r_{+}$ provided the mass
parameter $m$ lies between $m_{1,crit}\leq m\leq m_{2,crit} $ and
only an outer horizon if the parameter $m$ is outside of this
range, where $m_{1,crit}$ and $m_{2,crit}$ are the two critical
masses given by:
\begin{eqnarray}
m_{1,crit} &\equiv &-\frac l{3\sqrt{6}}\sqrt{1+33\frac{a^2}{l^2}(1-\frac{a^2%
}{l^2})-\frac{a^6}{l^6}+(1-14\frac{a^2}{l^2}+\frac{a^4}{l^4})^{3/2},}
\nonumber \\
m_{2,crit} &\equiv &-\frac l{3\sqrt{6}}\sqrt{1+33\frac{a^2}{l^2}(1-\frac{a^2%
}{l^2})-\frac{a^6}{l^6}-(1-14\frac{a^2}{l^2}+\frac{a^4}{l^4})^{3/2}.}
\label{mcrit}
\end{eqnarray}

The metric induced on the $2$-hypersurface at fixed $r
=$const$>r_{+}$ and $t=$const is Euclidean provided the parameter
$m\geq m_{1,crit}$. Thus although this metric has a horizon for
all values of $m$, to have a Euclidean metric on the $2$-surface
outside the horizon we should restrict the allowed values of the
mass parameter to $m\geq m_{1,crit}$. Note that this range
introduced here is a little greater than the one introduced in
Ref. \cite{Kl97} and this is due to the fact that we restrict
ourself to outside of the horizon.

Using Eq. (\ref{totact}) the action in the limit in which the
boundary becomes infinite can be written as
\begin{equation}
I_4=\beta _{+}^{(4)}\int_{\mathcal{B}_\infty }\mathcal{I}_B^{(4)}\sinh \chi
d\chi d\phi ,  \label{At4}
\end{equation}
where $\beta _{+}^{(4)}$ and $\mathcal{I}_B^{(4)}$ are the inverse
Hawking temperature of the event horizon and the action density on
the boundary at infinity given as
\begin{eqnarray}
\beta _{+}^{(4)} &=&\frac{4\pi l^2r_{+}(r_{+}^2+a^2)}{%
3r_{+}^4+(a^2-l^2)r_{+}^2+a^2l^2},  \label{bet4} \\
\mathcal{I}_B^{(4)} &=&-\frac 1{8\pi \Xi }\{m-r_{+}(\frac{r_{+}{}^2}{l^2}+3%
\frac{a^2}{l^2}\cosh ^2\chi )\}.  \label{Aden4}
\end{eqnarray}
Note that the $r$ divergences of the action density
$\mathcal{I}_B^{(4)}$ on
the boundary at infinity are removed, but since one cannot compactify $%
\mathcal{B}$, the divergences in $\chi $ still remain \cite{Kl98}.
Using Eqs. (\ref{Mas}) and (\ref{Ang}) the total mass $M$ and the
total angular momentum $J_\phi $ are

\begin{eqnarray}
M &=&\int_{\mathcal{B}_\infty }\mathcal{M}_B^{(4)}\sinh \chi d\chi d\phi ;%
\hspace{.5cm }\mathcal{M}_B^{(4)}=-\frac 1{4\pi }\frac m\Xi ,  \label{Mt4} \\
J_\phi &=&\int_{\mathcal{B}_\infty }\mathcal{J}_B^{(4)}\sinh \chi d\chi
d\phi ;\hspace{.5cm }\mathcal{J}_B^{(4)}=-\frac{3ma}{8\pi \Xi ^2}\sinh
^2\chi .  \label{Jt4}
\end{eqnarray}
These expressions show that the parameters $m$ and $a$ could be
associated with the mass and angular momentum of the spacetime,
respectively. As in the case of action, again the $r$ divergences
have been removed as one expects from AdS/CFT correspondence.

\subsection{Topological Kerr-AdS$_5$ Metric}

For the case of $n=4$ the metric given by Eqs. (\ref{met1a}) and
(\ref{met1b}) has two inner and outer horizons located at $r_{-}$
and $r_{+}$, provided the parameter $m$ lies between
$m_{1,crit}\leq m\leq m_{2,crit}$, and only an
outer horizon if the parameter $m>$ $m_{2,crit}$, where $m_{1,crit}$ and $%
m_{2,crit}$ are the two critical masses given by
\begin{eqnarray}
m_{1,crit} &\equiv &-\frac{l^2}8\left(1+\frac{a^2}{l^2}\right)^2,  \label{m1crit5} \\
m_{2,crit} &\equiv &-\frac 12a^2.  \label{m2crit5}
\end{eqnarray}
The metric induced on the $3$-hypersurface at fixed
$r=$const$>r_{+}$ and $t=$const is Euclidean for all the allowed
values of the mass parameter $m>m_{1,crit}$. Using Eq.
(\ref{totact}), one can write the total action as
\begin{equation}
I_5=\beta _{+}^{(5)}\int_{\mathcal{B}_\infty }\mathcal{I}_B^{(5)}\cosh \chi
\sinh \chi d\chi d\phi d\psi ,  \label{At5}
\end{equation}
where $\beta _{+}^{(5)}$ and $\mathcal{I}_B^{(5)}$ are the inverse
Hawking temperature of the event horizon and the action density on
the boundary at infinity is given as
\begin{eqnarray*}
\beta _{+}^{(5)} &=&\frac{2\pi l^2(r_{+}^2+a^2)}{r_{+}(2r_{+}^2+a^2-l^2)},
\label{bet5} \\
\mathcal{I}_B^{(5)} &=&-\frac 1{64\pi \Xi } \left\{8m+l^2 \left[3\left(1-\frac{a^2}{l^2}\right)^2-8%
\frac{r_{+}^4}{l^4}+2\left(9\frac{a^2}{l^2}-9\frac{a^4}{l^4}-8\frac{a^2r_{+}^2}{%
l^4}\right)\cosh ^2\chi +19\frac{a^4}{l^4}\cosh ^4\chi \right]
\right\}.
\end{eqnarray*}
Using Eqs. (\ref{Mas}) and (\ref{Ang}) the total mass $M$ and the
total angular momentum $J_\phi $ can be written as

\begin{eqnarray}
M &=&\int_{\mathcal{B}_\infty }\mathcal{M}_B^{(5)}\cosh \chi \sinh \chi
d\chi d\phi d\psi ,  \label{Mt5} \\
J_\phi  &=&\int_{\mathcal{B}_\infty }\mathcal{J}_B^{(5)}\cosh \chi \sinh
\chi d\chi d\phi d\psi ,  \label{Jt5}
\end{eqnarray}
where the mass and angular momentum densities $\mathcal{M}_B^{(5)}$ and $%
\mathcal{J}_B^{(5)}$ are
\begin{eqnarray}
\mathcal{M}_B^{(5)} &=&-\frac 1{64\pi \Xi }\left\{24m+l^2\left[(4-\Xi ^2)+2(6-\Xi )%
\frac{a^2}{l^2}\cosh ^2\chi +7\frac{a^4}{l^4}\cosh ^4\chi
\right]\right\},  \label{Mden5}
\\
\mathcal{J}_B^{(5)} &=&-\frac{ma}{2\pi \Xi ^2}\sinh ^2\chi .  \label{Jden5}
\end{eqnarray}
It is remarkable to note that the mass density
$\mathcal{M}_S^{(5)}$ computed in Eq. (\ref{Mden5}) is not zero in
the limit of $m=0$. This is a common feature for all the
asymptotically (A)dS spacetimes. Again the $r$ divergences of the
action, mass, and angular momentum densities are removed but the
divergences in $\chi $ still remain.

\subsection{Topological Kerr-AdS$_7$ Metric}

The topological Kerr-AdS$_7$ with one rotational parameter given
by Eqs. (\ref{met1a}) and (\ref{met1b}) has two inner and outer horizons located at $%
r_{-}$ and $r_{+}$ provided the mass parameter $m$ lies between $%
m_{1,crit}\leq m\leq m_{2,crit}$, and only an outer horizon if the parameter $%
m$ is greater than $m_{2,crit}$ where $m_{1,crit}$ and
$m_{2,crit}$ are the two critical masses given by
\begin{eqnarray}
m_{1,crit} &\equiv &-\frac{l^4}{27}\left\{1+\frac 32\frac{a^2}{l^2}\left(1-\frac{a^2}{%
l^2}\right)-\frac{a^6}{l^6}+\left(1+\frac{a^2}{l^2}+\frac{a^4}{l^4}\right)^{3/2}\right\},
\label{m1crit7} \\
m_{2,crit} &\equiv &-\frac{l^4}{27} \left\{1+\frac 32\frac{a^2}{l^2}\left(1-\frac{a^2}{%
l^2}\right)-\frac{a^6}{l^6}-\left(1+\frac{a^2}{l^2}+\frac{a^4}{l^4}\right)^{3/2}\right\}.
\label{M2crit7}
\end{eqnarray}
The metric induced on the $5$-hypersurface at fixed
$r=$const$>r_{+}$ and $t=$const is Euclidean for all the allowed
values of the mass parameter $m\geq m_{1,crit}$. Using Eq.
(\ref{totact}), one can write the total action as
\begin{equation}
I_7=\beta _{+}^{(5)}\int_{\mathcal{B}_\infty }\mathcal{I}_B^{(7)}\cosh
^3\chi \sinh \chi \cosh \psi \sinh \psi d\chi d\phi d\psi d\eta d\zeta ,
\label{Act7}
\end{equation}
where $\beta _{+}^{(7)}$ and $\mathcal{I}_B^{(7)}$ are the inverse
Hawking temperature of the event horizon and the action density on
the boundary at infinity given as
\begin{eqnarray*}
\beta _{+}^{(7)} &=&\frac{2\pi l^2r_{+}(r_{+}^2+a^2)}{%
3r_{+}^4+2r_{+}^2(a^2-l^2)-a^2l^2},  \label{Alph7} \\
\mathcal{I}_B^{(7)} &=&-\frac 1{640\pi \Xi } \{80m+l^4
[25(1-\frac{a^6}{l^6})-47 \frac{a^2}{l^2}(1-\frac{a^2}{l^2})-80
\frac{r^6_{+}}{l^6}\\ && +6(18 \Xi ^2-57 \frac{a^2}{l^2}-20
\frac{r^4_{+}}{l^4}) \frac{a^2}{l^2}\cosh ^2\chi
 +79 (1-\frac{a^2}{l^2}) \frac{a^4}{l^4}\cosh ^4
-20 \frac{a^6}{l^6}\cosh ^6\chi] \}.
\end{eqnarray*}
Again using Eqs. (\ref{Mas} \& \ref{Ang}), the total mass $M$ and the total
angular momentum $J_\phi $ can be written as:

\begin{eqnarray}
M &=&\int_{\mathcal{B}_\infty }\mathcal{M}_B^{(7)}\cosh ^3\chi \sinh \chi
\cosh \psi \sinh \psi d\chi d\phi d\psi d\eta d\zeta ;  \label{M7} \\
J_\phi  &=&\int_{\mathcal{B}_\infty }\mathcal{J}_B^{(7)}\cosh ^3\chi \sinh
\chi \cosh \psi \sinh \psi d\chi d\phi d\psi d\eta d\zeta ,  \label{J7}
\end{eqnarray}
where the mass and angular momentum densities are:
\begin{eqnarray*}
\mathcal{M}_B^{(7)} &=&-\frac 1{640\pi \Xi }\{400m+l^4[25-101\frac{a^2}{l^2}-%
\frac{a^4}{l^4}+5\frac{a^6}{l^6} \\
&&\ +3(63-56\frac{a^2}{l^2}-\frac{a^4}{l^4})\frac{a^2}{l^2}\cosh ^2\chi
+3(94-17\Xi )\frac{a^4}{l^4}\cosh ^4\chi +55\frac{a^6}{l^6}\cosh ^6\chi ]\},
\label{Mden7} \\
\mathcal{J}_B^{(7)} &=&-\frac 3{4\pi }\frac{ma}{\Xi ^2}\sinh ^2\chi .
\label{Jden7}
\end{eqnarray*}
Although the divergences in $r$ have been removed, the $\chi $
divergences still remain.

\section{The General Topological Kerr-AdS Metric in Five Dimension \label
{GenK5}}

The general rotating black hole in five dimensions can have at
most two rotational parameters \cite{Hawk}. Thus one can write the
metric of five dimensional topological black brane with two
rotational parameters as
\begin{eqnarray}
ds^2 &=&-\frac{\Delta _r}{\rho ^2}(dt+\frac{a\sinh ^2\theta }{\Xi _a}d\phi -%
\frac{b\cosh ^2\theta }{\Xi _a}d\psi )^2-\frac{\rho ^2}{\Delta _\theta }%
d\theta ^2+\frac{\rho ^2}{\Delta _r}dr^2  \nonumber \\
&&\ \ \ \ \ \ +\frac{1+r^2/l^2}{r^2\rho ^2}[a b
dt+\frac{b(r^2+a^2)\sinh ^2\theta }{\Xi _a}d\phi
-\frac{a(r^2+b^2)\cosh ^2\theta }{\Xi _b}d\psi ]^2
\nonumber \\
&&\ \ \ \ \ \ -\frac{\Delta _\theta \sinh ^2\theta }{\rho ^2}(a dt-\frac{%
r^2+a^2}{\Xi _a}d\phi )+\frac{\Delta _\theta \cosh ^2\theta }{\rho ^2}(b dt-%
\frac{r^2+b^2}{\Xi _b}d\phi ),  \label{metr2a}
\end{eqnarray}
where
\begin{eqnarray}
\Delta _r &=&\frac 1{r^2}(r^2+a^2)(r^2+b^2)(1+\frac{r^2}{l^2})-2m,  \nonumber
\\
\Delta _\theta  &=&1-\frac{a^2}{l^2}\cosh ^2\theta +\frac{b^2}{l^2}\sinh
^2\theta ,  \nonumber \\
\Xi _a &=&1+\frac{a^2}{l^2},\hspace{.5cm}\Xi _b=1+\frac{b^2}{l^2},  \nonumber
\\
\rho ^2 &=&r^2+a^2\cosh ^2\theta -b^2\sinh ^2\theta .  \label{metr2b}
\end{eqnarray}
The metric given by Eqs. (\ref{metr2a}) and (\ref{metr2b}) is the
solution of the Einstein equation with negative cosmological
constant. It has two inner and outer
horizons provided the mass parameter $m$ is greater than the critical mass $%
m_{crit}$:
\begin{eqnarray}
m_{crit} &=&\frac 1{24l^2}\{\Upsilon ^{1/3}+6^4\mu ^2(a^2b^2l^2)^{4/3}(1+\mu
^6)\Upsilon ^{-1/3}+15\mu ^4(a^2b^2l^2)^{2/3}-4(a^4+b^4+l^4)\} ,  \nonumber \\
\Upsilon  &=&18^3a^4b^4l^4\{1+20\mu ^6-\mu ^{12}+(1-8\mu
^6)^{3/2}\} , \nonumber \\
\mu^6&=&\frac{(a^2+b^2+l^2)^3}{216a^2b^2l^2}. \label{Mcr5b}
\end{eqnarray}
Using Eq. (\ref{totact}), one can write the total action as
\begin{equation}
I_5^{\prime }=\beta _{+}^{(5)}\int_{\mathcal{B}_\infty }\mathcal{I}%
_S^{\prime (5)}\cosh \chi \sinh \chi d\chi d\phi d\psi ,  \label{Act5b}
\end{equation}
where $\beta _{+}^{\prime (5)}$ and $\mathcal{I}_B^{\prime (5)}$
are the inverse Hawking temperature of the event horizon and the
action density on the boundary at infinity given as
\begin{eqnarray*}
\beta _{+}^{\prime (5)} &=&-\frac{2\pi l^2r_{+}(r_{+}^2+a^2)(r_{+}^2+b^2)}{%
2r_{+}^6+(a^2+b^2+l^2)r^4+a^2b^2l^2},  \label{A25b} \\
\mathcal{I}_B^{\prime (5)} &=&-\frac 1{64\pi \Xi _a\Xi _b}\{8m+3(a^2+l^2)(1-4%
\frac{b^2}{l^2})+\frac 4{l^2}(b^4-b^2r_{+}^2-2r_{+}^4) \\
&&\ \ -2[10\Xi _b-9\Xi _a+8(1+\frac{r_{+}^2}{l^2})](a^2-b^2)\cosh ^2\chi +%
\frac{19}{l^2}(a^2-b^2)^2\cosh ^4\chi \}.
\end{eqnarray*}
Again the total mass $M$ and the total angular momenta $J_\phi $
and $J_\psi $ calculated from Eqs. (\ref{Mas}) and (\ref{Ang}) are
given by Eqs. (\ref{Mt5}) and (\ref{Jt5}) where the mass and
angular momentum densities are

\begin{eqnarray*}
\mathcal{M}_B^{\prime (5)} &=&-\frac 1{64\pi \Xi _a\Xi _b}\{24m+l^2(4\Xi
_b^2-\Xi _a^2) \\
&&\ \ -2(6\Xi _b-\Xi _a)(a^2-b^2)\cosh ^2\chi +\frac 7{l^2}(a^2-b^2)^2\cosh
^4\chi \},  \label{M5b} \\
\mathcal{J}_\phi ^{\prime (5)} &=&-\frac{ma}{2\pi \Xi _a^2\Xi
_b}\sinh ^2\chi ,\hspace{.5cm }\mathcal{J}_\psi ^{\prime
(5)}=-\frac{mb}{2\pi \Xi _a\Xi _b^2}\cosh ^2\chi .  \label{J5b}
\end{eqnarray*}
Again note that the mass density in the limit of $m=0$ vanishes and all the $%
r$ divergences in the densities are removed.

\section{Closing Remarks}

In this paper we have considered a class of higher genus solutions
of the Einstein equation with negative cosmological constant which
represents rotating topological black branes with one rotating
parameter in more than three dimensions. Also we have introduced
the general topological Kerr-AdS solutions in five dimensions with
two rotating parameters. Since one is interested in outside the
horizon we have modified the allowed values of the mass parameter
$m$, for topological Kerr-AdS$_4$ given in Ref. \cite{Kl97}, and
introduced these values for topological Kerr-AdS in various
dimensions up to seven.

The main aim of this paper was the investigation of the efficiency of the
AdS/CFT correspondence for these rotating topological black branes. We have
found out that the counterterms introduced in \cite{Kls} cannot remove the $%
r $ divergences of the action of these spacetimes in more than
five dimensions. But as it is well known, the expression for
$I_{ct}$ obtained by the algorithm given in ref. \cite{Kls} is not
unique, and one can add any functional of curvature invariants
that vanish at infinity in a given dimension. Thus in order to
remove the $r$ divergences in action we modified these
counterterms by adding a suitable curvature invariant term to it.
Although this term vanishes for black holes whose horizons have
negative or positive curvature constant, we have shown that it
removes the divergences of the action of higher genus rotating
topological black branes in more than five dimensions. Using
modified counterterms, we have computed the conserved mass and
locally conserved angular momentum densities of rotating
topological Kerr-AdS spacetimes in various dimensions through the
use of a Brown-York boundary stress tensor. We found out that
although the $r$ divergences of all the densities on the boundary
$\mathcal{B}_\infty $ are removed, the divergences in $\chi $
still remain. This is due to the fact that one cannot compactify
the boundary $\mathcal{B}$. As in the case of asymptotically (A)dS
black holes, we found out that in the limit of $m=0$ the mass
density of these spacetimes in odd dimensions is not zero.

The efficiency of the AdS/CFT correspondence for higher
dimensional rotating topological black branes and other solutions
of the Einstein equation such as cosmological $C$-metric and the
extension of these ideas to the case of asymptotically de Sitter
spacetimes remain interesting subjects for future investigations.
Also removing the divergences in $\chi $ of the action and
conserved charges and calculating finite values of these
quantities remain an open issue, which we leave for the future.

\end{document}